\documentclass[twocolumn,prb,showpacs]{revtex4}
\usepackage{graphicx}
\usepackage{dcolumn}
\usepackage{bm}
\usepackage{amsmath}

\begin{document}

\preprint{}
\title{ Anisotropic supercurrent due to inhomogeneous magnetization in ferromagnet/superconductor junctions}
\author{Takehito Yokoyama$^1$ Yukio Tanaka$^2$, and Shuichi Murakami$^1$}
\affiliation{$^1$Department of Physics, Tokyo Institute of Technology, Tokyo 152-8551,
Japan 
\\ $^2$Department of Applied Physics, Nagoya University, Nagoya, 464-8603, Japan
}
\date{\today}

\begin{abstract}
We investigate transverse charge and spin dc supercurrents in a ferromagnet coupled to a superconductor where the ferromagnet has inhomogeneous magnetic structure. These transverse supercurrents arise from non-trivial structure of the magnetization. The magnetic structure manifested in the transverse charge supercurrent is essentially different from that discussed in the context of anomalous Hall effect, reflecting the disspationless nature of supercurrent. Possible candidates of magnetic structure to verify our prediction are also discussed.

\end{abstract}

\maketitle


\section{Introduction}

The interplay between superconductivity and ferromagnetism has received much attention.\cite{Buzdin,Bergeret,Wang,Eschrig,Takahashi,Blamire,Linder2,Eschrig2,Beckmann,Sidorenko} In particular, generation of spin-triplet pairing in ferromagnet/superconductor junction  is of paramount importance.\cite{Bergeret2} Equal-spin triplet pairing emerges due to spin flip scattering in ferromagnetic multilayer or inhomogeneous ferromagnet. Spin-polarized supercurrent, carried by equal-spin triplet pairing,  is a new ingredient for spintronics applications. Several experiments have successfully demonstrated the presence of spin-triplet pairing by observing Josephson current through strong ferromagnet.\cite{Keizer,Khaire,Robinson}  
In ferromagnetic Josephson junctions, longitudinal Josephson current has been investigated\cite{Eschrig2,Bergeret3,Golubov,Eschrig4,Asano,Eschrig3,Alidoust}. More recently, transverse (Hall) Josephson current has been predicted in ferromagnetic Josephson junctions which stems from triplet superconductivity\cite{Yokoyama} or spin orbit couplings\cite{Linder3,Malshukov,Costa}.

The Hall effect in ferromagnet has been discussed intensively in the context of anomalous Hall effect. \cite{Nagaosa}
The anomalous Hall effect arises from non-trivial spin structure, which is associated with the spin Berry phase effect. \cite{Ye,Taguchi,Ohgushi,Tatara,Onoda}
It is shown that the Hall conductivity contains the terms steming from non-trivial spin configurations such as vector spin chirality ${\bf{S}}_i  \times {\bf{S}}_j$ \cite{Taguchi2} and scalar spin chirality  ${\bf{S}}_i  \cdot ({\bf{S}}_j  \times {\bf{S}}_k )$ \cite{Tatara}, where ${\bf{S}}_i$ is a localized spin with position $i$. Non-trivial spin structures also give rise to dissipationless spin current\cite{Chandra,Konig,Lee,Nogueira,Bruno,Takeuchi,Sonin,Chen}. Motivated by these studies, in this paper, we consider transverse supercurrent driven by non-trivial magnetic structure under phase gradient. Since the phase is odd in time-reversal, the magnetic structure manifested in transverse supercurrent becomes essentially different from that in the anomalous Hall effect.

In this paper, we study transverse charge and spin dc supercurrents in a ferromagnet coupled to a superconductor where the ferromagnet has inhomogeneous magnetic structure. Analytic expressions of the transverse super currents are obtained based on perturbative calculation. 
The transverse supercurrents arise from non-trivial structure of the magnetization. The magnetic structure manifested in the transverse charge supercurrent is essentially different from that discussed in the context of anomalous Hall effect, reflecting the disspationless nature of supercurrent. Possible candidates of magnetic structure to verify our prediction are discussed.

\section{Formulation}

\begin{figure}[tbp]
\begin{center}
\scalebox{0.8}{
\includegraphics[width=8.50cm,clip]{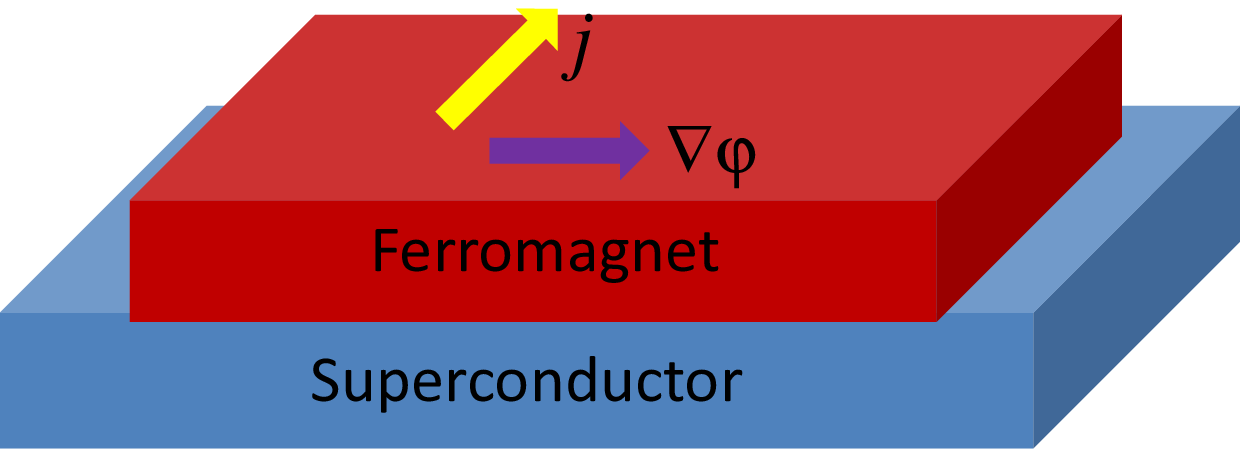}
}
\end{center}
\caption{ Schematic illustration of the ferromagnet/superconductor junction.}
\label{fig3}
\end{figure}

We consider a ferromagnet/superconductor junction (See  Fig. \ref{fig3}).
The Hamiltonian of the superconductor and the ferromagnet are given by  $H_S  = H_0  + H_\Delta $ and $H_F  = H_0  + H_{ex} + H_\varphi$, respectively. 
The $H_0$, $H_\Delta$ and  $H_{ex}$ represent the kinetic energy, the superconducting order, and the exchange interaction between the conducting electron and the local spins, respectively:
\begin{eqnarray}
 H_0  = \sum\limits_{\bf{k}} {\phi _{\bf{k}}^\dag  \xi \tau _3 \phi _{\bf{k}}^{} },  \\ 
 H_\Delta   = \sum\limits_{\bf{k}} {\phi _{\bf{k}}^\dag  \Delta \tau _2 \phi _{\bf{k}}^{} },  \\ 
 H_{ex}  =  - J\sum\limits_{{\bf{k}},{\bf{q}}} {(\phi _{{\bf{k}} - {\bf{q}}}^\dag  {\bm{\sigma }}\phi _{\bf{k}}^{} ) \cdot {\bf{n}}_{\bf{q}}^{} }  \label{hex}
\end{eqnarray}
with $\xi  = \varepsilon _k  - \varepsilon _F  \equiv \frac{{\hbar ^2 k^2 }}{{2m}} - \varepsilon _F $ and $\phi _{\bf{k}}^\dag   = (c_{{\bf{k}} \uparrow }^\dag  ,c_{{\bf{k}} \downarrow }^\dag  ,ic_{ - {\bf{k}} \downarrow }^{} , - ic_{ - {\bf{k}} \uparrow }^{} )$  where  $\sigma$ and $\tau$ are Pauli matrices in spin and Nambu spaces, respectively. $\varepsilon _F$, $\Delta$, $J$, and $\bf{n}$ are the Fermi energy, the gap function, the exchange coupling, and the unit vector pointing in the direction of the local spins, respectively. 
The localized spins can have spatial dependence, but we consider only slowly varying case compared to the Fermi wavelength. 
Note that we adopt the basis in Ref.\cite{Ivanov} such that singlet pairing is proportional to the unit matrix in spin space. 
We consider supercurrent induced by phase gradient.
The phase gradient along $j$ direction, $\nabla _j \varphi$, enters the Hamiltonian as 
\begin{eqnarray}
H_\varphi   = \sum\limits_{\bf{k}} {\phi _{\bf{k}}^\dag  \frac{{\hbar^2}}{m}k_j \nabla _j \varphi \phi _{\bf{k}}^{} } 
\end{eqnarray}
where
$\nabla _j \varphi$ is assumed to be spatially constant. 
We will treat $H_{ex}$ and $H_\varphi$ perturbatively. 

With the above Hamiltonians,
the charge ($j_{c}$) and spin ($j_{s}$) current operators in $i$-direction read \begin{eqnarray}
j_{c,i}^{}  =  - \frac{{e\hbar }}{m}k_i  - \delta _{ij} \frac{{e\hbar  }}{m}\nabla _j \varphi \tau _3 , \\
j_{s,i}^\alpha   = \frac{{\hbar ^2 }}{{2m}}k_i \tau _3 \sigma ^\alpha   + \delta _{ij} \frac{{\hbar^2}}{{2m}}\nabla _j \varphi \sigma ^\alpha
\end{eqnarray}
where $-e$ is the electron charge and $\alpha$ denotes the direction of spin.

\section{Results}

Before proceeding to the explicit calculation, let us discuss transverse supercurrents qualitatively based on the time-reversal symmetry. \cite{Murakami}
Consider the London equation, 
\begin{equation}
{\bf{j}}_c  =  - \frac{{e^2 }}{m}\rho  \cdot {\bf{A}},
\end{equation}
where ${\bf{j}}_c$, $\rho$, and ${\bf{A}}$ are, respectively,  the charge current, the superfluid density tensor, and  the vector potential. 
Since the charge current and the vector potential are time-reversal odd, $\rho$ describes the reversible and dissipationless flow of the supercurrent. Thus, the transverse current can flow without breaking time-reversal symmetry. Namely, the transverse current is allowed in even-order perturbation with respect to time-reversal breaking term $H_{ex}$. 
This contrasts with the anomalous Hall effect\cite{Tatara} where the Hall current is driven by the electric field which is even under time-reversal. Thus, one can expect essentially different magnetic structure manifested in the transverse supercurrent. 
Similarly, let us consider response equation of spin current, 
\begin{equation}
{\bf{j}}_s  =   \frac{{\hbar e }}{2m}\rho'  \cdot {\bf{A}}.
\end{equation}
where ${\bf{j}}_s$ and $\rho'$ are the spin current and the superfluid density tensor for spin current, respectively. 
Since spin current is even under time-reversal, $\rho'$ relates quantities of different symmetries under time-reversal. Thus, the time-reversal symmetry should be broken to produce finite spin current within the linear response. 
Since $\rho'$ contains time-reversal breaking perturbation  $H_{ex}$, this argument indicates that spin current appears only in odd-order perturbation with respect to the exchange interaction.

\begin{figure}[tbp]
\begin{center}
\scalebox{0.8}{
\includegraphics[width=9.50cm,clip]{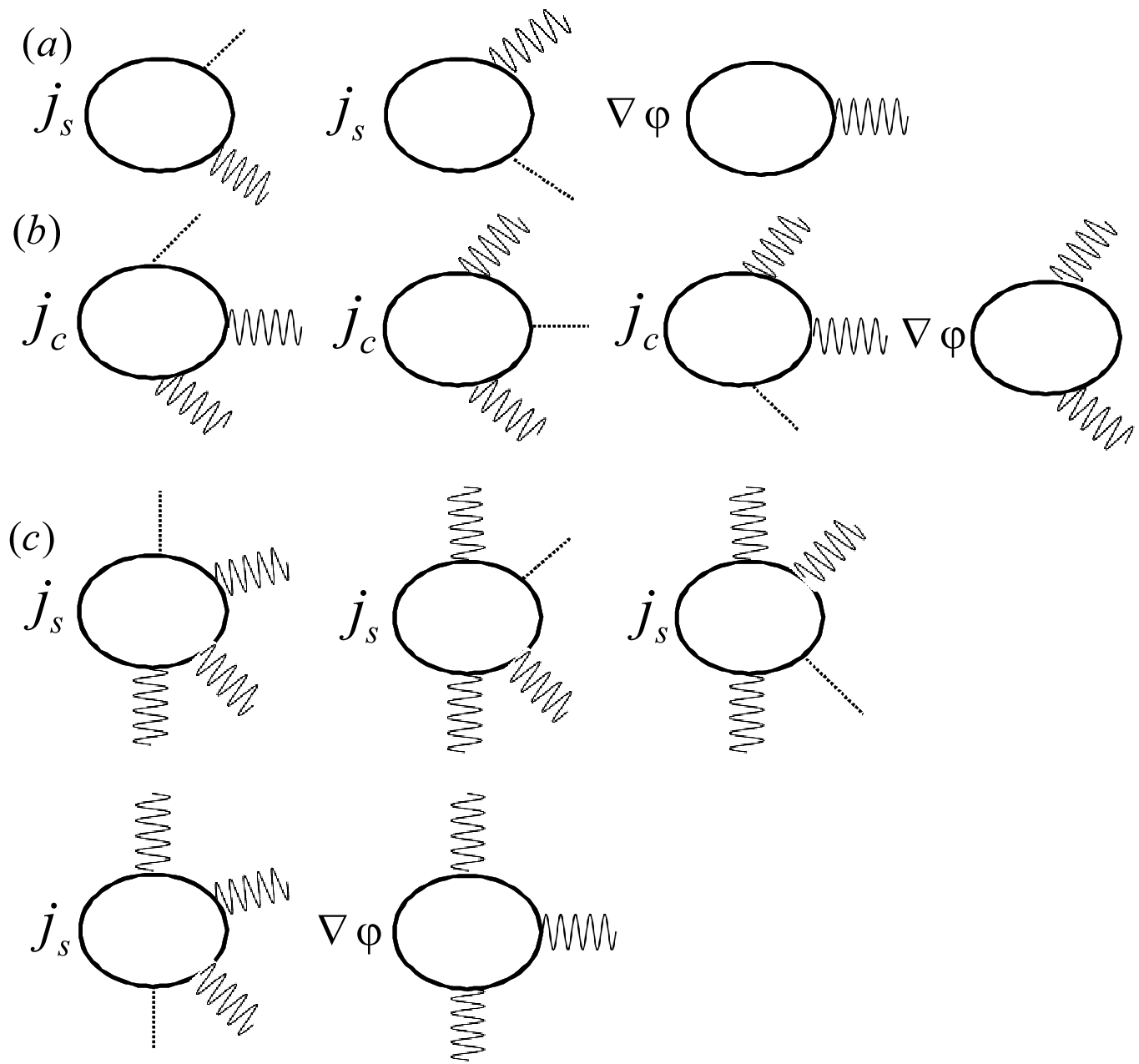}
}
\end{center}
\caption{ Diagrammatic representations of the current densities. Diagrams (a) describe first order contributions in $J$, (b) second-order contributions and (c) third-order contributions. The wavy lines denote the interaction with the local spin $\bf{n}$ and dotted lines represent the phase gradient $\nabla \varphi$. The last diagrams in (a), (b), and (c) correspond to the second terms in Eqs.(5) and (6).}
\label{fig1}
\end{figure}


Now, we calculate transverse supercurrents and give their analytical expressions. Our central results are Eqs.(10), (11), (13) and (14). We consider the unperturbed advanced scalar Green's functions in the ferromagnet of the form 
$g_{{\bf{k}},\omega }^a  = g_{0,{\bf{k}},\omega }^a  + g_{3,{\bf{k}},\omega }^a \tau _3  + f_{{\bf{k}},\omega }^a \tau _2$
where $g_{0,{\bf{k}},\omega }^a$ and $g_{3,{\bf{k}},\omega }^a$ are normal Green's functions while $f_{{\bf{k}},\omega }^a $ is the anomalous Green's function. The anomalous Green's function in the ferromagnet arises due to the proximity effect.
We take into account $H_{ex}$ up to the third order and $H_\varphi$ as a first order perturbation. Diagrammatic representations of the transverse currents are shown in Fig. \ref{fig1}.
We first consider transverse charge supercurrent which can be represented as \cite{Haug}
\begin{eqnarray}
 j_{c,i}  =  \frac{{i\hbar ^2 e}}{{mV}}\sum\limits_{{\bf{k}},{\bf{q}}} {e^{ - i{\bf{q}} \cdot {\bf{x}}} {\rm{Tr}}k_i G_{{\bf{k}} - {\bf{q}}/2,{\bf{k}} + {\bf{q}}/2}^ <  (t,t)} \nonumber \\ 
  + \delta _{ij} \frac{{i\hbar^2  e}}{{mV}}\nabla _j \varphi \sum\limits_{{\bf{k}},{\bf{q}}} {e^{ - i{\bf{q}} \cdot {\bf{x}}} {\rm{Tr}}\tau _3 G_{{\bf{k}} - {\bf{q}}/2,{\bf{k}} + {\bf{q}}/2}^ <  (t,t)} 
\end{eqnarray}
where $V$ is the total volume and ${\rm{Tr}}$ is taken over spin and Nambu spaces. $G_{{\bf{k}} - {\bf{q}}/2,{\bf{k}} + {\bf{q}}/2}^ <  (t,t)$ is the lesser Green's function of the total Hamiltonian. 
Performing perturbation with respect to $H_{ex}$ and $H_\varphi$, we expand the lesser component using the advanced Green's functions by the Langreth theorem.\cite{Haug}
Noting that $g_{{\bf{k}},\omega }^ <   = f_\omega  \left[ {g_{{\bf{k}},\omega }^a  - (g_{{\bf{k}},\omega }^a )^\dag  } \right]$ with the lesser Green's function $g_{{\bf{k}},\omega }^ <$  and the Fermi distribution function $f_\omega$, and $\delta _{ij}  = \frac{{\partial k_i }}{{\partial k_j }}$, we can compute the transverse charge and spin supercurrents  (See Appendix for details).
 The first order expansion in $J$ vanishes since the Green's function is proportional to the unit matrix in the spin space. 
Then, the leading  term of the transverse charge current ($i \ne j$) is in the second order in $J$. The current reads 
\begin{eqnarray}
{j_{c,i}} \cong  - \frac{{e\hbar }}{m}\rho _{ij}^{}{\nabla _j}\varphi,
\end{eqnarray}
\begin{widetext}
\begin{eqnarray}
\rho _{ij}^{} = {\delta _{ij}}\frac{{512}}{3}{J^2}\sum\limits_{{\bf{k}},\omega } {{f_\omega }{\mathop{\rm Im}\nolimits} {{(f_{{\bf{k}},\omega }^a)}^2}\left\{ {5{{(g_{0,{\bf{k}},\omega }^a)}^2} + {{(f_{{\bf{k}},\omega }^a)}^2} + {{(g_{3,{\bf{k}},\omega }^a)}^2}} \right\}} \nonumber \\
 - {\delta _{ij}}\left( {{\nabla ^2}{\bf{n}}(x) \cdot {\bf{n}}(x)} \right)\frac{{64{\hbar ^3}}}{{9Vm}}{J^2}\sum\limits_{{\bf{k}},\omega } {{f_\omega }{\mathop{\rm Im}\nolimits} }  \times \nonumber \\
\left[ {\varepsilon _k^2{{(f_{{\bf{k}},\omega }^a)}^2}\left\{ {15{{(g_{0,{\bf{k}},\omega }^a)}^4} - 2{{(g_{0,{\bf{k}},\omega }^a)}^2}\left\{ {7{{(f_{{\bf{k}},\omega }^a)}^2} - 33{{(g_{3,{\bf{k}},\omega }^a)}^2}} \right\} - {{\left\{ {{{(f_{{\bf{k}},\omega }^a)}^2} + {{(g_{3,{\bf{k}},\omega }^a)}^2}} \right\}}^2}} \right\} + 24\varepsilon _k^{}{{(f_{{\bf{k}},\omega }^a)}^2}{{(g_{0,{\bf{k}},\omega }^a)}^2}g_{3,{\bf{k}},\omega }^a} \right] \nonumber\\
 + \left( {{\nabla _i}{\bf{n}}(x) \cdot {\nabla _j}{\bf{n}}(x)} \right)\frac{{128{\hbar ^3}}}{{9Vm}}{J^2}\sum\limits_{{\bf{k}},\omega } {{f_\omega }{\mathop{\rm Im}\nolimits} }  \times \nonumber \\
\left[ {\varepsilon _k^2{{(f_{{\bf{k}},\omega }^a)}^2}\left\{ {15{{(g_{0,{\bf{k}},\omega }^a)}^4} - 2{{(g_{0,{\bf{k}},\omega }^a)}^2}\left\{ {7{{(f_{{\bf{k}},\omega }^a)}^2} - 33{{(g_{3,{\bf{k}},\omega }^a)}^2}} \right\} - {{\left\{ {{{(f_{{\bf{k}},\omega }^a)}^2} + {{(g_{3,{\bf{k}},\omega }^a)}^2}} \right\}}^2}} \right\} + 12\varepsilon _k^{}{{(f_{{\bf{k}},\omega }^a)}^2}{{(g_{0,{\bf{k}},\omega }^a)}^2}g_{3,{\bf{k}},\omega }^a} \right].
\end{eqnarray}
\end{widetext}
If the anomalous Green's function $f_{{\bf{k}},\omega }^a$ becomes zero, then $\rho =0$ as expected. We have also found by the explicit calculation that the third order perturbation with respect to $J$ does not contribute to the transverse current.  
Thus, up to the third order in $J$, only second order perturbation with respect to $J$ remains finite as expected from the above argument based on the time-reversal symmetry.
It is also seen from Eq.(10) that the superfluid density tensor is symmetric: $\rho _{ij}=\rho _{ji}$. Therefore, there is no Hall effect in our setup.

Next, we will calculate transverse spin supercurrent. 
The spin current  is calculated as 
\begin{eqnarray}
j_{s,i}^\alpha   =  - \frac{{i\hbar ^3 }}{{2mV}}\sum\limits_{{\bf{k}},{\bf{q}}} {e^{ - i{\bf{q}} \cdot {\bf{x}}} {\rm{Tr}}k_i \tau _3 \sigma ^\alpha  G_{{\bf{k}} - {\bf{q}}/2,{\bf{k}} + {\bf{q}}/2}^ <  (t,t)} \nonumber \\ 
  - \delta _{ij} \frac{{i\hbar ^3}}{{2mV}}\nabla_j  \varphi\sum\limits_{{\bf{k}}, {\bf{q}}} {e^{ - i{\bf{q}} \cdot {\bf{x}}} {\rm{Tr}}\sigma ^\alpha  G_{{\bf{k}} - {\bf{q}}/2,{\bf{k}} + {\bf{q}}/2}^ <  (t,t)}.
\end{eqnarray}
In the first order in $J$, the spin current is represented as
\begin{widetext}
\begin{eqnarray}
j_{s,i}^\alpha  \cong \frac{{{\hbar ^2}}}{{2m}}\rho _{ij}^{\prime}{\nabla _j}\varphi \\\rho _{ij}^{\prime} = {\delta _{ij}}\frac{{32{\hbar ^3}}}{{3Vm}}\nabla _{}^2{\bf{n}}^\alpha ({\bf{x}}) \sum\limits_{{\bf{k}},\omega } {{f_\omega }{\mathop{\rm Im}\nolimits} } \left[ {{\varepsilon _k}\left( {1 + \frac{8}{3}{\varepsilon _k}g_{3,{\bf{k}},\omega }^a} \right){{(f_{{\bf{k}},\omega }^a)}^2}\left\{ { - {{(g_{0,{\bf{k}},\omega }^a)}^2} + {{(f_{{\bf{k}},\omega }^a)}^2} + {{(g_{3,{\bf{k}},\omega }^a)}^2}} \right\}} \right] \nonumber \\
 + \frac{{32{\hbar ^3}}}{{3Vm}}\nabla _i^{}\nabla _j^{}{\bf{n}}^\alpha ({\bf{x}}) \sum\limits_{{\bf{k}},\omega } {{f_\omega }{\mathop{\rm Im}\nolimits} } \left[ {{\varepsilon _k}\left( {1 + \frac{{8(1+\delta_{ij})}}{3}{\varepsilon _k}g_{3,{\bf{k}},\omega }^a} \right){{(f_{{\bf{k}},\omega }^a)}^2}\left\{ { - {{(g_{0,{\bf{k}},\omega }^a)}^2} + {{(f_{{\bf{k}},\omega }^a)}^2} + {{(g_{3,{\bf{k}},\omega }^a)}^2}} \right\}} \right].
\end{eqnarray}
\end{widetext}
It is seen that when the anomalous Green's function $f_{{\bf{k}},\omega }^a$ becomes zero, then $\rho ^{\prime}=0$. By the explicit calculation, we also find that the second order term with respect to $J$ vanishes, which is consistent with the above argument based on the time-reversal symmetry.  
The third order expansion with respect to $J$ yields finite contribution to the transverse spin current. The detailed expression is quite complicated and hence omitted here. The transverse spin current in the third order in $J$ has the form,
\begin{eqnarray}
j_{s,i}^\alpha   = J^3 [ A'\nabla _i \nabla _j {\bf{n}}_{}^\alpha  ({\bf{x}}) \nonumber \\
+ B'(\nabla _i {\bf{n}}({\bf{x}}) \cdot \nabla _j {\bf{n}}({\bf{x}})){\bf{n}}_{}^\alpha  ({\bf{x}}) ] \nabla _j \varphi 
\end{eqnarray}
wherer $A'$ and $B'$ depend solely on junction parameters.
Also, we find that the superfluid density tensor for spin current is symmetric: $\rho' _{ij}=\rho' _{ji}$. Hence, there is no Hall effect for spin supercurrent.

Therefore, under the phase gradient in $x$-direction, up to the third order in $J$, we have the transverse charge and spin supercurrents in $y$-direction driven by magnetic structure of the form:
\begin{eqnarray}
 j_{c,y}  =  - \frac{{e \hbar}}{m} J^2 \rho _c^{} \left( {\partial _x {\bf{n}}({\bf{x}}) \cdot \partial _y {\bf{n}}({\bf{x}})} \right)\nabla _x \varphi, \label{jc}  \\ 
  j_{s,y}^\alpha   =  [(\frac{{\hbar^2}}{{2m}}  J \rho _s + J^3 A')\partial _x \partial _y {\bf{n}}_{}^\alpha  ({\bf{x}}) \nonumber \\ 
  + J^3 B'(\partial _x {\bf{n}}({\bf{x}}) \cdot \partial _y {\bf{n}}({\bf{x}})){\bf{n}}_{}^\alpha  ({\bf{x}})] \nabla _x \varphi. \label{js}
\end{eqnarray}
These structures contrast with the normal Hall current in the ferromagnet:
In the normal state, the Hall current is driven by scalar spin chirality under electric field \cite{Tatara}
\begin{eqnarray}
j_{c,y}  \propto \left( {\frac{\partial }{{\partial x}}{\bf{n}}({\bf{x}}) \times \frac{\partial }{{\partial y}}{\bf{n}}({\bf{x}})} \right) \cdot {\bf{n}}({\bf{x}}). \label{jcn}
\end{eqnarray}
Equilibrium spin current driven by inhomogeneous magnetic structure in the normal state is given by \cite{Takeuchi}
\begin{eqnarray}
 j_{s,y}^\alpha \propto  \left( {\frac{\partial }{{\partial y}}{\bf{n}}({\bf{x}}) \times {\bf{n}}({\bf{x}})} \right)^\alpha. \label{jsn}
\end{eqnarray}
By comparing Eq.(\ref{jc}) and Eq.(\ref{js}), and, Eq.(\ref{jcn}) and Eq.(\ref{jsn}), we find essentially different magnetic structures required for transverse supercurrents, which reflects the fact that supercurrent flows in response to phase gradient, the disspationless nature of supercurrent.

\begin{figure}[tbp]
\begin{center}
\scalebox{0.8}{
\includegraphics[width=8.0cm,clip]{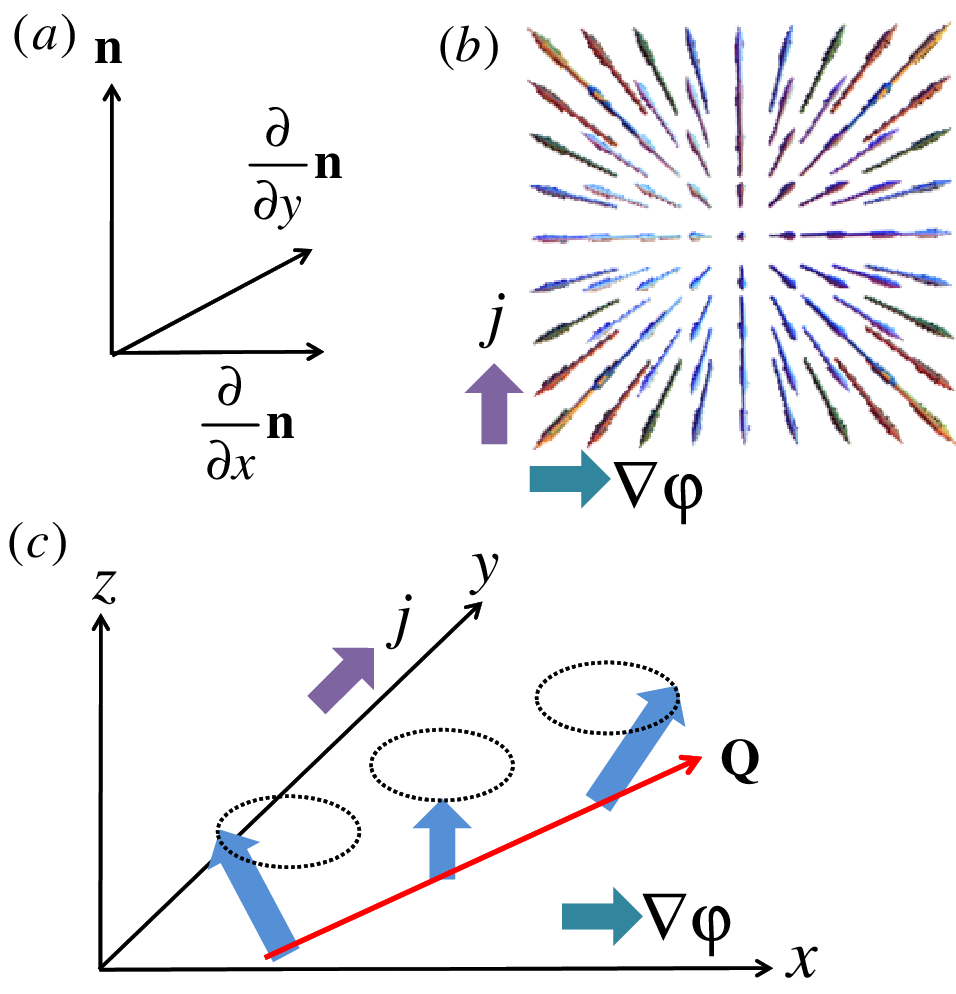}}
\end{center}
\caption{(Color online) (a) Magnetization vector $\bf{n}$. (b) Vortex spin structure. (c) Conical spin structure. The transverse supercurrents under the phase gradient arise in these spin structures.}
\label{fig2}
\end{figure}

Now, we discuss possible candidates of magnetic structure to verify our prediction. First, the magnetization vector ${\bf{n}}(\bf{x})$ should have both $x$ and $y$ dependence. To observe the transverse charge supereffect, $\partial _x {\bf{n}}(\bf{x})$ and $\partial _y {\bf{n}}(\bf{x})$, both perpendicular to ${\bf{n}}(\bf{x})$, should not be perpendicular to each other (see Fig. \ref{fig2} (a)). 
One possible candidate is a spin vortex structure (or magnetic skyrmions in chiral magnets\cite{Rossler,Muhlbauer,Nagaosa2,Fert}) as shown in Fig. \ref{fig2} (b) where ${\bf{n}}({\bf{x}})$ is characterized by 
${\bf{n}}({\bf{x}}) = \frac{1}{a}(x,y,\sqrt {a^2  - x^2  - y^2 } )$ with a real constant $a$. Then, we have 
\begin{eqnarray}
\frac{\partial }{{\partial x}}{\bf{n}}({\bf{x}}) \cdot \frac{\partial }{{\partial y}}{\bf{n}}({\bf{x}}) =   \frac{{xy}}{{a^2 (a^2  - x^2  - y^2) }}, \\ 
\frac{{\partial ^2 }}{{\partial x\partial y}}{\bf{n}}({\bf{x}}) = \frac{1}{a}(0,0,\frac{{-xy}}{{(a^2  - x^2  - y^2 )^{3/2} }}).
\end{eqnarray}
For $xy \ne 0$, we obtain nonzero transverse supercurrents.
A conical ferromagnet, as illustrated Fig. \ref{fig2} (c), is another candidate ferromagnet. The magnetization vector ${\bf{n}}({\bf{x}})$ can be written as  
${\bf{n}}({\bf{x}}) = \frac{1}{{\sqrt {1 + b^2 } }}(\cos ({\bf{Q}} \cdot {\bf{x}}),\sin ({\bf{Q}} \cdot {\bf{x}}),b)$
where $\bf{Q}$ is a magnetic vector and $b$ is a real constant. Then, we have 
\begin{eqnarray}
 \frac{\partial }{{\partial x}}{\bf{n}}({\bf{x}}) \cdot \frac{\partial }{{\partial y}}{\bf{n}}({\bf{x}}) = \frac{{Q_x Q_y }}{{1 + b^2 }},  \label{conical1} 
\end{eqnarray}
\begin{eqnarray}
\frac{{\partial ^2 }}{{\partial x\partial y}}{\bf{n}}({\bf{x}}) =  - Q_x Q_y \frac{1}{{\sqrt {1 + b^2 } }}(\cos ({\bf{Q}} \cdot {\bf{x}}),\sin ({\bf{Q}} \cdot {\bf{x}}),0).
\end{eqnarray}
Therefore, for $Q_x Q_y  \ne 0$, we obtain finite transverse super currents.

Since the Josephson junction composed of a conical ferromagnet Ho has been fabricated,\cite{Robinson}  our prediction could be confirmed by junctions with this material in four-terminal geometry. 
The presence of the predicted transverse spin current could be probed experimentally by conversion into an electrical signal via the inverse spin Hall effect (by injecting the spin current into a spin-orbit coupled normal metal)\cite{Saitoh,Valenzuela}.
The transverse currents reflect a local magnetic texture. Local measurement of these currents can be used to determine the magnetization profile.

When the proximity effect is strong such that the Green's functions in the ferromagnet have the same form as those in the bulk superconductor:
\begin{eqnarray}
g_{{\bf{k}},\omega }^a  = \frac{{\omega  - i\gamma  + \xi \tau _3  + \Delta \tau _2 }}{{(\omega  - i\gamma )^2  - \xi ^2  - \Delta ^2 }}
\end{eqnarray}
where $\gamma$ is the inelastic scattering rate by impurities, the transverse charge current Eq. (\ref{jc}) can be reduced to 
\begin{eqnarray}
j_{c,y}  \cong 0.06 \times \frac{{64e\hbar ^3 }}{{9m^2}}\frac{{\nu \varepsilon _F^2 }}{{\Delta ^4 }}J^2 \left( {{\partial _x } {\bf{n}}({\bf{x}}) \cdot {\partial _y} {\bf{n}}({\bf{x}})} \right)\nabla _x \varphi  \label{jcb} 
\end{eqnarray}
for $\gamma  \ll \Delta$ at zero temperature where $\nu$ is the density of states at the Fermi level. 
Let us estimate the transverse current for conical ferromagnet using Eqs. (\ref{conical1}) and (\ref{jcb}). 
For $\varepsilon _F \sim 1$ eV, $J \sim $ 1 meV, $b = 1/\tan (4\pi /9) \cong 40$, $\nabla _x \varphi  \sim (100 $ nm$)^{ - 1}, Q_x  \cong Q_y  \sim (1$ nm$)^{- 1}$, $\nu \sim 0.1$ /eV/unit cell, $\Delta \sim$ 1 meV, and  the lattice constant $\sim 5$ \AA,  we estimate the magnitude of the current as $j_{c,y} \sim 3 \times 10^{8}$ A/cm$^2$.

Spin Hall effect due to the Rashba-type spin-orbit coupling in superconductors~\cite{Kontani} or Josephson junctions~\cite{Mal'shukov} has been discussed.
In this paper, we have predicted transverse supercurrent driven by non-trivial magnetic structure, and hence our results do not rely on spin-orbit coupling. 
In Ref. ~\cite{Mal'shukov}, spin Hall effect is obtained by applying electric bias to the Josephson junction in order to make the current time-dependent. 
In stark contrast, we have considered stationary supercurrent under non-trivial magnetic structure when a phase gradient is applied.


Generation of dissipationless spin current has been also predicted in non-trivial spin structures such as interfaces between two ferromagnets in the normal states \cite{Chandra,Konig,Lee,Nogueira,Bruno,Takeuchi,Sonin,Chen}. 
The mechanism proposed in this paper is completely different: It requires both a gradient in spin space and a phase gradient, and the resulting spin current is carried by Cooper pairs.

\section{Conclusions}

In summary,
 we have studied transverse charge and spin supercurrents in a ferromagnet coupled to a superconductor where the ferromagnet has inhomogeneous magnetization. The transverse supercurrents stem from non-trivial structure of the magnetization. The magnetic structure manifested in the transverse charge supercurrent is essentially different from that discussed in the context of anomalous Hall effect, reflecting the disspationless nature of supercurrent. 

This work was supported by JSPS KAKENHI Grant
Number JP30578216 and
Scientific Research (A) (KAKENHI Grant No. JP20H00131 and JP18H03678), Scientific Research (B) (KAKENHI Grant Numbers JP18H01176 and JP20H01857), Japan RFBR Bilateral Joint Research Projects/Seminars number 19-52-50026, and the JSPS Core-to-Core program "Oxide Superspin" international network.

\section*{Appendix}

\begin{widetext}

Here, we present some details of the calculations of charge and spin supercurrents. We focus on the off-diagonal components  ($i \ne j$). Diagonal components can be calculated in a similar way. 
The charge supercurrent corresponding to Fig. 2(b) reads
\begin{eqnarray}
{j_{c,i}} = \frac{{2i{\hbar ^2}e}}{{mV}}\frac{{{\hbar ^2}}}{m}{J^2}{\sum\limits_{{\bf{k}},{\bf{q}},{\bf{q}}',\omega} {{e^{ - i{\bf{q}} \cdot {\bf{x}}}}{\nabla _j}\varphi {{\bf{n}}_{{\bf{q}}'}} \cdot {{\bf{n}}_{{\bf{q}} - {{\bf{q}}^\prime }}}{\mathop{\rm Tr}\nolimits} {{\bf{k}}_i}\left[ \begin{array}{l}
{g_{{\bf{k}} - {\bf{q}}/2,\omega}}{({\bf{k}} - {\bf{q}}/2)_j}{g_{{\bf{k}} - {\bf{q}}/2,\omega}}{g_{{\bf{k}} - {\bf{q}}/2 + {\bf{q}}',\omega}}{g_{{\bf{k}} + {\bf{q}}/2,\omega}}\\
 + {g_{{\bf{k}} - {\bf{q}}/2,\omega}}{g_{{\bf{k - q}}/2 + {\bf{q}}',\omega}}{({\bf{k}} - {\bf{q}}/2 + {\bf{q}}')_j}{g_{{\bf{k}} - {\bf{q}}/2 + {\bf{q}}',\omega}}{g_{{\bf{k}} + {\bf{q}}/2,\omega}}\\
 + {g_{{\bf{k}} - {\bf{q}}/2,\omega}}{g_{{\bf{k}} - {\bf{q}}/2 + {\bf{q}}',\omega}}{g_{{\bf{k}} + {\bf{q}}/2,\omega}}{({\bf{k}} + {\bf{q}}/2)_j}{g_{{\bf{k}} + {\bf{q}}/2,\omega}}
\end{array} \right]} ^ < } \nonumber\\ 
 + \delta_{ij} \frac{{i{\hbar ^2}e}}{{mV}}{J^2}{\nabla _j}\varphi \sum\limits_{{\bf{k}},{\bf{q}},{\bf{q}}',\omega} {{e^{ - i{\bf{q}} \cdot {\bf{x}}}}{{\bf{n}}_{{\bf{q}}'}} \cdot {{\bf{n}}_{{\bf{q}} - {{\bf{q}}^\prime }}}{\mathop{\rm Tr}\nolimits} {\tau _3}{{\left[ {{g_{{\bf{k}} - {\bf{q}}/2,\omega}}{g_{{\bf{k}} - {\bf{q}}/2 + {\bf{q}}',\omega}}{g_{{\bf{k}} + {\bf{q}}/2,\omega}}} \right]}^ < }} .
\end{eqnarray}
We expand the lesser component using the advanced Green's functions by the Langreth theorem.\cite{Haug}
Noting that $g_{{\bf{k}},\omega }^ <   = f_\omega  \left[ {g_{{\bf{k}},\omega }^a  - (g_{{\bf{k}},\omega }^a )^\dag  } \right]$ with the lesser Green's function $g_{{\bf{k}},\omega }^ <$ and $\delta _{ij}  = \frac{{\partial k_i }}{{\partial k_j }}$, and expanding the Green's functions up to the second order of spatial gradient of the local spins, we obtain
\begin{eqnarray}
j_{c,i} =  - \frac{{4{\hbar ^2}e{J^2}}}{{mV}}\sum\limits_{{\bf{k}},{\bf{q}},{\bf{q}}',\omega} {{e^{ - i{\bf{q}} \cdot {\bf{x}}}}{\nabla _j}\varphi {{\bf{n}}_{{\bf{q}}'}} \cdot {{\bf{n}}_{{\bf{q}} - {{\bf{q}}^\prime }}}{f_\omega }} \nonumber \\
 \times {\mathop{\rm Im}\nolimits} {\mathop{\rm Tr}\nolimits} \frac{{{\hbar ^2}}}{m}{{\bf{k}}_i}\left[ \begin{array}{l}
2{{\bf{k}}_i}{\bf{k}}_j^2{{\bf{q}}_i}'{{\bf{q}}_j}'{\left( {\frac{{{\hbar ^2}}}{m}} \right)^2}\{ 4{(g_{{\bf{k}},\omega }^a)^5}{\tau _3}g_{{\bf{k}},\omega }^a{\tau _3} + {(g_{{\bf{k}},\omega }^a)^4}{\tau _3}{(g_{{\bf{k}},\omega }^a)^2}{\tau _3}\\
 - 2{\tau _3}g_{{\bf{k}},\omega }^a{\tau _3}{(g_{{\bf{k}},\omega }^a)^2}{\tau _3}g_{{\bf{k}},\omega }^a{\tau _3}{(g_{{\bf{k}},\omega }^a)^2} - 3{\tau _3}{(g_{{\bf{k}},\omega }^a)^2}{\tau _3}g_{{\bf{k}},\omega }^a{\tau _3}g_{{\bf{k}},\omega }^a{\tau _3}{(g_{{\bf{k}},\omega }^a)^2}\} 
\end{array} \right.\nonumber \\
\left. { + 2\frac{{{\hbar ^2}}}{m}{{\bf{k}}_i}{{\bf{q}}_i}'{{\bf{q}}_j}'\left\{ {{{(g_{{\bf{k}},\omega }^a)}^5}{\tau _3} - {\tau _3}{{(g_{{\bf{k}},\omega }^a)}^2}{\tau _3}g_{{\bf{k}},\omega }^a{\tau _3}{{(g_{{\bf{k}},\omega }^a)}^2}} \right\}} \right]
\end{eqnarray}
which reduces to Eqs.(10) and (11) by taking the trace. 

The spin supercurrent corresponding to Fig. 2(a) reads
\begin{eqnarray}
j_{s,i}^\alpha  =  - \frac{{i{\hbar ^5}J}}{{{m^2}V}}{\nabla _j}{\varphi }{\sum\limits_{{\bf{k}},{\bf{q}},\omega} {{e^{ - i{\bf{q}} \cdot {\bf{x}}}}{\bf{n}}_{\bf{q}}^\alpha {\mathop{\rm Tr}\nolimits} {{\bf{k}}_i}{\tau _3}\left[ {{g_{{\bf{k}} - {\bf{q}}/2,\omega}}{{({\bf{k}} - {\bf{q}}/2)}_j}{g_{{\bf{k}} - {\bf{q}}/2,\omega}}{g_{{\bf{k}} + {\bf{q}}/2,\omega}} + {g_{{\bf{k}} - {\bf{q}}/2,\omega}}{g_{{\bf{k}} + {\bf{q}}/2}}{{({\bf{k}} + {\bf{q}}/2)}_j}{g_{{\bf{k}} + {\bf{q}}/2,\omega}}} \right]} ^ < }\nonumber \\
 - \delta_{ij} \frac{{i{\hbar ^5}J}}{{{m^2}V}}{\nabla _j}\varphi \sum\limits_{{\bf{k}},{\bf{q}},\omega} {{e^{ - i{\bf{q}} \cdot {\bf{x}}}}{\bf{n}}_{\bf{q}}^\alpha {\mathop{\rm Tr}\nolimits} {{\left[ {{g_{{\bf{k}} - {\bf{q}}/2,\omega}}{g_{{\bf{k}} + {\bf{q}}/2,\omega}}} \right]}^ < }} .
\end{eqnarray}
In a way similar to the charge supercurrent, we have 
\begin{eqnarray}
j_{s,i}^\alpha  = \frac{{2{\hbar ^5}J}}{{{m^2}V}}{\nabla _j}\varphi \sum\limits_{{\bf{k}},{\bf{q}},\omega } {{e^{ - i{\bf{q}} \cdot {\bf{x}}}}{\bf{n}}_{\bf{q}}^\alpha {f_\omega }{\rm Im} {\mathop{\rm Tr}\nolimits} \left[ \begin{array}{l}
2{\bf{k}}_i^2{\bf{k}}_j^2{{\bf{q}}_i}{{\bf{q}}_j}{\left( {\frac{{{\hbar ^2}}}{m}} \right)^2}\left( {{\tau _3}g_{{\bf{k}},\omega }^a{\tau _3}g_{{\bf{k}},\omega }^a{\tau _3}{{(g_{{\bf{k}},\omega }^a)}^3} - {\tau _3}{{(g_{{\bf{k}},\omega }^a)}^2}{\tau _3}g_{{\bf{k}},\omega }^a{\tau _3}{{(g_{{\bf{k}},\omega }^a)}^2}} \right)\\
 + {\bf{k}}_i^2{{\bf{q}}_i}{{\bf{q}}_j}\frac{{{\hbar ^2}}}{m}\left( {{\tau _3}g_{{\bf{k}},\omega }^a{\tau _3}{{(g_{{\bf{k}},\omega }^a)}^3} - {\tau _3}{{(g_{{\bf{k}},\omega }^a)}^2}{\tau _3}{{(g_{{\bf{k}},\omega }^a)}^2}} \right)
\end{array} \right]}
\end{eqnarray}
which reduces to Eqs.(13) and (14) by taking the trace.

\end{widetext}


\begin{thebibliography}{99}

\bibitem{Buzdin} A. I. Buzdin, Rev. Mod. Phys. \textbf{77}, 935 (2005).

\bibitem{Bergeret} F. S. Bergeret, A. F. Volkov, and K. B. Efetov, Rev. Mod. Phys. \textbf{77}, 1321 (2005).

\bibitem{Wang} J. Wang, M. Singh, M. Tian, N. Kumar, B. Liu, C. Shi, J. K. Jain, N. Samarth, T. E. Mallouk, and M. H. W. Chan, 
Nat. Phys. \textbf{6}, 389 (2010). 

\bibitem{Eschrig} M. Eschrig, Physics Today \textbf{64}, 43 (2011). 

\bibitem{Takahashi} S. Takahashi and S. Maekawa, J. Phys. Soc. Jpn. \textbf{77}, 031009 (2008).

\bibitem{Blamire} M. G. Blamire and J. W. A. Robinson, J. Phys.: Condens. Matter \textbf{26}, 453201 (2014).

\bibitem{Linder2} J. Linder and J. W. A. Robinson, Nat. Phys. \textbf{11}, 307 (2015).

\bibitem{Eschrig2} M. Eschrig, Rep. Prog. Phys. \textbf{78}, 104501 (2015).

\bibitem{Beckmann} D. Beckmann, J. Phys.: Condens. Matter \textbf{28}, 163001 (2016).

\bibitem{Sidorenko} A. Sidorenko (Ed.), \textit{Functional Nanostructures and Metamaterials for Superconducting Spintronics} (Springer, 2018).


\bibitem{Bergeret2} F. S. Bergeret, A. F. Volkov, and K. B. Efetov, Phys. Rev. Lett. \textbf{86}, 4096 (2001); Phys. Rev. B \textbf{64}, 134506 (2001).

\bibitem{Keizer} R. S. Keizer, S. T. B. Goennenwein, T. M. Klapwijk, G. Miao, G. Xiao, and A. Gupta, 
Nature (London) \textbf{439}, 825 (2006).

\bibitem{Khaire} T. S. Khaire, M. A. Khasawneh, W. P. Pratt, Jr., and N. O. Birge, Phys. Rev. Lett. \textbf{104}, 137002 (2010).

\bibitem{Robinson} J. W. A. Robinson, J. D. S. Witt, M. G. Blamire, Science \textbf{329}, 59 (2010) 


\bibitem{Bergeret3} F. S. Bergeret, A. F. Volkov, and K. B. Efetov, Phys. Rev. B \textbf{68}, 064513 (2003); A. F. Volkov, F. S. Bergeret, and K. B. Efetov, Phys. Rev. Lett. \textbf{90}, 117006 (2003).

\bibitem{Golubov} A. A. Golubov, M. Yu. Kupriyanov, and E. llichev, Rev. Mod. Phys. \textbf{76}, 411 (2004).

\bibitem{Eschrig4} 
M. Eschrig, J. Kopu, J. C. Cuevas, and G. Schon, Phys. Rev. Lett. \textbf{90}, 137003 (2003).

\bibitem{Asano} 
Y. Asano, Y. Tanaka, and A. A. Golubov, Phys. Rev. Lett. \textbf{98}, 107002 (2007);
Y. Asano, Y. Sawa, Y. Tanaka, and A. A. Golubov, Phys. Rev. B \textbf{76}, 224525 (2007).

\bibitem{Eschrig3} 
M. Eschrig and T. L\"{o}fwander, Nature Phys. \textbf{4}, 138 (2008).

\bibitem{Alidoust} 
M. Alidoust, J. Linder, G. Rashedi, T. Yokoyama, and A. Sudbo, Phys. Rev. B \textbf{81}, 014512 (2010).

\bibitem{Yokoyama} T. Yokoyama, Phys. Rev. B \textbf{92}, 174513 (2015).

\bibitem{Linder3} J. Linder, M. Amundsen, and V. Risingg\r{a}rd, Phys. Rev. B \textbf{96}, 094512 (2017).

\bibitem{Malshukov} A. G. Mal'shukov, Phys. Rev. B \textbf{100} 035301 (2019).

\bibitem{Costa} A. Costa and J. Fabian, Phys. Rev. B \textbf{101} 104508 (2020).


\bibitem{Nagaosa} N. Nagaosa, J. Sinova, S. Onoda, A. H. MacDonald, and P. Ong,  Rev. Mod. Phys. \textbf{82}, 1539 (2010). 

\bibitem{Ye} J. Ye, Y. B. Kim, A. J. Millis, B. I. Shraiman, P. Majumdar, and Z. Tesanovic, 
Phys. Rev. Lett. \textbf{83}, 3737 (1999).

\bibitem{Taguchi} Y. Taguchi, Y. Oohara, H. Yoshizawa, N. Nagaosa, and Y. Tokura,  Science \textbf{291}, 2573 (2001).

\bibitem{Ohgushi} K. Ohgushi, S. Murakami, and N. Nagaosa, Phys. Rev. B \textbf{62}, R6065 (2000).

\bibitem{Tatara} G. Tatara and H. Kawamura, J. Phys. Soc. Jpn. \textbf{71}, 2613 (2002).

\bibitem{Onoda} M. Onoda, G. Tatara, and N. Nagaosa, J. Phys. Soc. Jpn. \textbf{73}, 2624 (2004).

\bibitem{Taguchi2} K. Taguchi and G. Tatara, Phys. Rev. B \textbf{79}, 054423 (2009).

\bibitem{Chandra} P. Chandra, P. Coleman, and A. I. Larkin, J. Phys. Condens. Matter \textbf{2}, 7933 (1990).

\bibitem{Konig} J. K\"{o}nig, M. Chr. B${\o}$nsager, and A. H. MacDonald, Phys. Rev. Lett. \textbf{87}, 187202 (2001).

\bibitem{Lee} Y.-L. Lee and Y.-W. Lee, Phys. Rev. B \textbf{68}, 184413 (2003).

\bibitem{Nogueira} F. S. Nogueira and K.-H. Bennemann, Europhys. Lett. \textbf{67}, 620 (2004).

\bibitem{Bruno} P. Bruno and V. K. Dugaev, Phys. Rev. B \textbf{72}, 241302(R) (2005).

\bibitem{Takeuchi} A. Takeuchi and G. Tatara, J. Phys. Soc. Jpn. \textbf{77}, 074701 (2008); A. Takeuchi, K. Hosono, and G. Tatara, Phys. Rev. B \textbf{81}, 144405 (2010).

\bibitem{Sonin} E. B. Sonin, Adv. Phys. \textbf{59}, 181 (2010).

\bibitem{Chen} W. Chen, P. Horsch, and D. Manske, Phys. Rev. B \textbf{89}, 064427 (2014).

\bibitem{Ivanov} D. A. Ivanov and Ya. V. Fominov, Phys. Rev. B \textbf{73}, 214524 (2006).

\bibitem{Murakami} S. Murakami, N. Nagaosa and Shou-cheng Zhang, Phys. Rev. B \textbf{69}, 235206 (2004).


\bibitem{Haug} H. Haug and A.-P. Jauho, \textit{Quantum Kinetics in Transport and Optics of Semiconductors} (Springer, New York, 1997).

\bibitem{Rossler} U. K. R\"ossler, A. N. Bogdanov, and C. Pfleiderer, Nature (London) \textbf{442}, 797 (2006).

\bibitem{Muhlbauer} S. M\"uhlbauer, B. Binz, F. Jonietz, C. Pfleiderer, A. Rosch, A. Neubauer, R. Georgii, and P. B\"oni, Science \textbf{323}, 915 (2009).

\bibitem{Nagaosa2} N. Nagaosa and Y. Tokura, Nat. Nanotechnol. \textbf{8}, 899 (2013).

\bibitem{Fert} A. Fert, V. Cros, and J. Sampaio, Nat. Nanotechnol. \textbf{8}, 152 (2013).

\bibitem{Saitoh}  E. Saitoh, M. Ueda, H. Miyajima, and G. Tatara, 
Appl. Phys. Lett. \textbf{88}, 182509 (2006).

\bibitem{Valenzuela} S. O. Valenzuela and M. Tinkham, Nature (London) \textbf{442}, 176 (2006).

\bibitem{Kontani} H. Kontani, J. Goryo, and D. S. Hirashima, Phys. Rev. Lett. \textbf{102}, 086602 (2009).

\bibitem{Mal'shukov} A. G. Mal'shukov and C. S. Chu, Phys. Rev. B \textbf{78}, 104503 (2008); Phys. Rev. B \textbf{84}, 054520 (2011).



\end{thebibliography}
\end{document}